# Crystal orientation relation and macroscopic surface roughness in hetero-epitaxially grown graphene on Cu/mica

J. L. Qi[1, 2]*, K. Nagashio[2+], T. Nishimura[2] and A. Toriumi[2]

[1] *State Key Laboratory of Advanced Welding and Joining, Department of Materials Science and Engineering, Harbin Institute of Technology, Harbin 150001 China*

[2] *Department of Materials Engineering, The University of Tokyo, 7-3-1 Hongo, Bunkyo, Tokyo 113-8656 Japan*

Corresponding authors:
*E-mail: jlqi@hit.edu.cn (J. L. Qi)
+E-mail: nagashio@material.t.u-tokyo.ac.jp (K. Nagashio)

**Abstract:** A clean, flat and orientation-identified graphene on a substrate is in high demand for graphene electronics. In this study, the hetero-epitaxial graphene growth on Cu(111)/mica(001) by chemical vapor deposition is investigated to check the applicability for the top-gate insulator research on graphene as well as the graphene channel research by transferring graphene on $SiO_2$/Si substrates. After adjusting the graphene-growth condition, the surface roughness of the graphene/Cu/mica substrate and the average smooth area are ~0.34 nm and ~100 μm$^2$, respectively. The orientation of graphene in the graphene/Cu/mica substrate can be identified by the hexagonal void morphology of Cu. Moreover, we demonstrate the relatively high mobility of ~4500 cm$^2$V$^{-1}$s$^{-1}$ in graphene transferred on the $SiO_2$/Si substrate. These results suggest that the present graphene/Cu/mica substrate can be used for top-gate insulator research on graphene.

## 1. Introduction

Practical applications of graphene electronic devices require the establishment of an electrically reliable top-gate insulator on graphene [1-4]. However, the progress in the insulator research on graphene is slow in spite of its importance, because a clean, flat and orientation-identified graphene on substrates is not widely available, compared with conventional Si substrates. For example, the small grain size in highly-oriented pyrolytic graphite and the limited area of graphene available by mechanical exfoliation prevent the systematic and reliable study for physical, chemical and electrical characterizations of the top-gate insulator on graphene.

The conditions required for top-gate insulator research on graphene are as follows: (i) large area single-crystalline graphene, (ii) low surface roughness in the relatively large area (RMS < ~0.5 nn for ~100 μm□), (iii) negligible defects in graphene, (iv) resist-residue-free and wrinkle-free surface of graphene, (v) a well-identified crystal orientation of graphene. Namely, a catalytic metal under graphene is not necessary to be removed for the insulator research on graphene. Thus, we define this system as the graphene substrate.

So far, using the chemical vapor deposition (CVD) technique, the single-crystalline graphene is successfully grown on Cu(111) grown hetero-epitaxially on sapphire substrate [5-7] and mica substrate [8] or on bulk single crystal Ni(111) [9] substrate. Mica seems to be most likely the best of the aforementioned substrate choices because the flat Cu(111) surface is attainable on the atomically smooth surface of mica. Moreover, mica can be reused by peeling off the Cu-deposited mica surface and do not require high-temperature surface treatment at 1500 °C to obtain atomically flat surfaces, compared with sapphire. Although it can be expected to use the graphene/Cu/mica sample as the graphene substrate, detailed information on crystal orientation relations in graphene/Cu(111)/mica(001) and macroscopic surface roughness have not been reported.

Concerning the crystalline quality of graphene grown by CVD, the heavy evaporation of Cu at low-pressure conditions has been shown to affect the nucleation and subsequent growth of graphene and its electrical properties [10-13]. Therefore, the key to further improving the quality of graphene is to reduce both agglomeration and evaporation of Cu during the CVD growth.

In this paper, the hetero-epitaxial graphene growth on Cu(111)/mica(001) by chemical vapor deposition is investigated to assess the applicability as the graphene substrate from the viewpoint of surface roughness and crystal orientation relations in graphene/Cu(111)/mica(001). Furthermore, we show the crystalline quality improvement of graphene by the CVD growth at ambient pressure.

## 2. Experimental procedure

The atomically flat surface of single-crystal mica (001) ($KMg_3(AlSi_3O_{10})F_2$) substrate (10×10×0.5 $mm^3$) was simply obtained by cleaving with the tape. Cu films with a thickness of ~800 nm were hetero-epitaxially deposited on the mica substrate heated to 500 °C. The film thicker than ~600 nm was required to avoid the agglomeration and void formation during the CVD growth at 1000 °C.

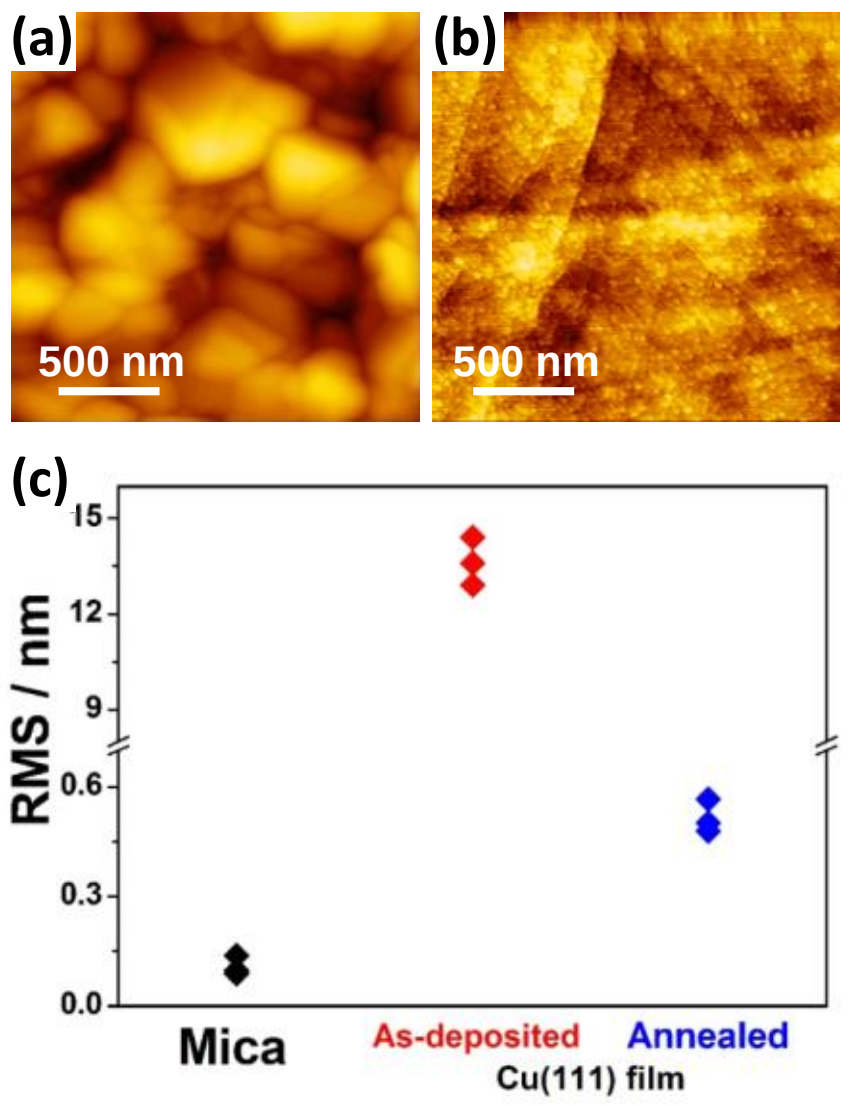

**Fig. 1** AFM images of (a) the as-deposited Cu surface on the mica substrate; (b) the Cu surface on the mica substrate after the annealing in $H_2$/Ar gas flow (100/50 sccm) at 1000 °C for 30 min; and (c) RMS roughness for mica, as-deposited Cu/mica and annealed Cu/mica surfaces, respectively.

Before the graphene growth, the Cu/mica substrate was annealed in an $H_2$/Ar gas flow (100/50 sccm) at 1000 °C for 30 min, in a three-zone tube furnace with the uniform temperature range (±1 °C) of ~ 10 cm. Subsequently, graphene was grown on the Cu/mica substrate by additionally introducing $CH_4$ in the chamber at the same temperature. The total pressure in the chamber was kept roughly at 1000 Pa or at ambient pressure. The crystallographic characterization was performed using X-ray diffraction (XRD) and electron backscattering pattern (EBSP). The surface roughness (RMS) was analyzed by atomic force microscopy (AFM).

The Cu film on mica was dissolved in HCl aqueous solution, and graphene was then transferred onto $SiO_2$ (90 nm)/Si substrate using the polymethyl methacrylate (PMMA)-assisted transfer method [14]. The transferred graphene was analyzed by the Raman microscope with Ar laser unit of 488 nm excitation. Moreover, field effect transistors of single-layer graphene (SLG) were fabricated using conventional electron beam (EB) lithography techniques, and their electrical properties were characterized in vacuum at room temperature [15].

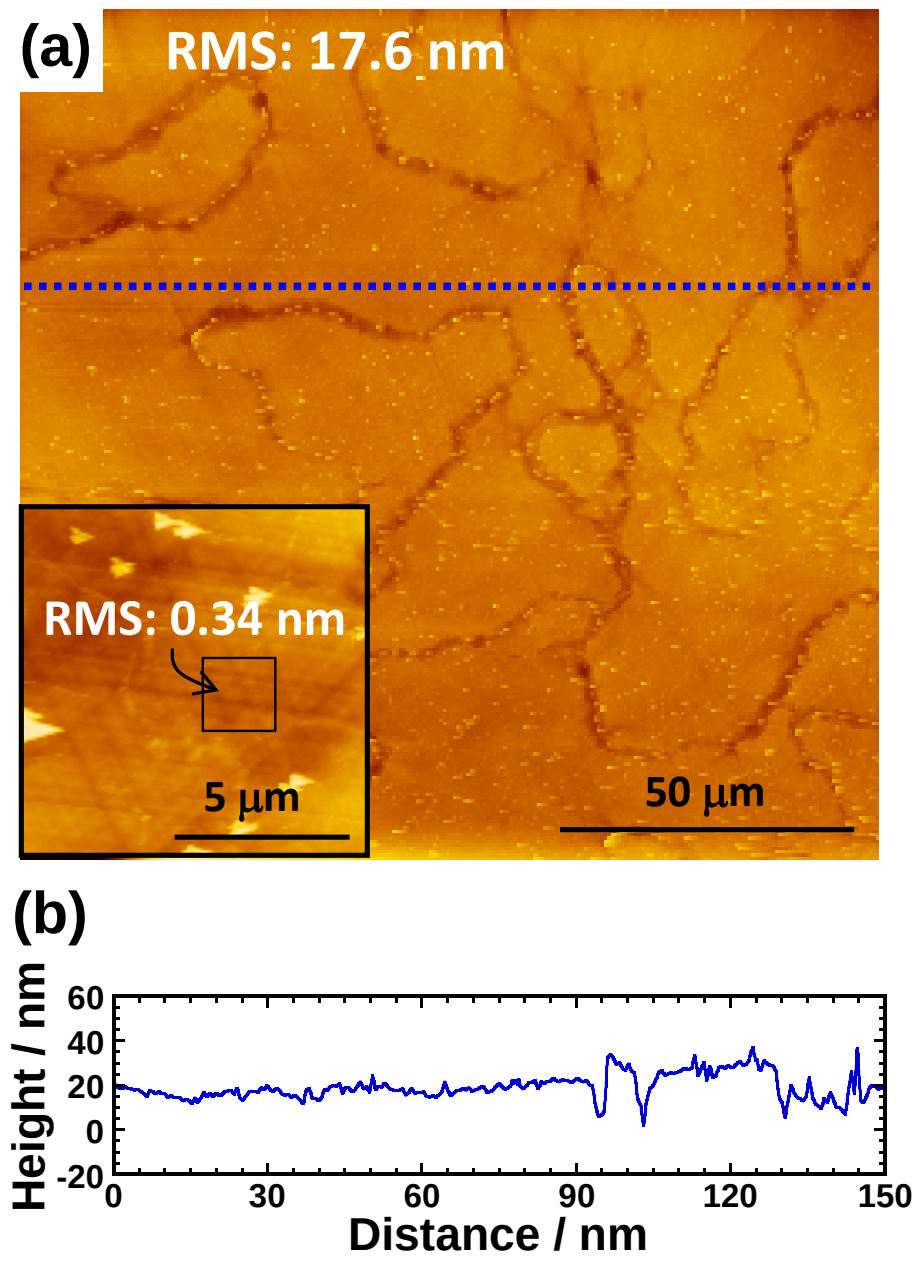

**Fig. 2** (a) The AFM image for graphene on the Cu/mica substrate with relatively large area (150 $\mu m^2$) and (b) height profile along the dotted line in (a). The inset in (a) shows the AFM image of the small area.

## 3. Results & Discussion

### 3-1. Characterization of Graphene/Cu/mica substrate

**Figure 1** shows the AFM images for (a) as-deposited Cu surface and (b) Cu surface after the annealing in Ar/$H_2$ gas flow at 1000 °C for 30 min. XRD analysis indicated that the Cu film was single crystal oriented to (111). Heating the mica substrate up to 500 °C during Cu deposition is critical for the hetero-epitaxial growth. The triangular shape, which is characteristic of the (111) close-packed plane for fcc metals [16], can clearly be observed after the annealing of Cu/mica substrate. RMS was reduced from 13.6 nm to 0.5 nm after the annealing, as shown in **Fig. 1(c)**.

Subsequently, SLG was grown on Cu/mica. The detailed growth conditions are described in section 3.2. **Figure 2** shows (a) the AFM image for graphene on the Cu/mica substrate with a relatively large area (150 $\mu m^2$) and (b) the height profile along the dotted line in (a). The inset in (a) indicates that the local area surface of the graphene/Cu/mica substrate is smoother than that of the Cu/mica substrate in **Fig. 1(b)** because the Cu surface covered by graphene is not oxidized [17]. The triangular shape is also clearly evident in the inset of **Fig. 2(a)**. On the other hand, the grooves, as indicated by black arrows, can be observed in the macroscopic image of (a) because of the twin formation as a result of relaxing the lattice mismatch. The average smooth area is ~100 $\mu m^2$, which is not large enough but is acceptable for the conventional electrode size in capacitance-voltage measurement when the graphene/Cu/mica substrate is used as the graphene substrate.

To identify the orientation of graphene on Cu/mica, the voids with the hexagonal shape in the Cu film were focused. **Figure 3(a)** shows the optical micrographs at the edge and center of the Cu/mica substrate (red circles in the inset). This sample was annealed at 1000 °C intentionally for longer than 1 hour to promote the void formation in the Cu film. The voids appear preferentially along the step edge of mica. The orientation of the hexagonal shape for voids is consistent throughout the Cu/mica substrate, as indicated by the red dotted lines, suggesting that the Cu film is macroscopically single crystal. **Figure 3(b)** shows the EBSP map of the Cu film around the hexagonal void. The transverse direction is colored in

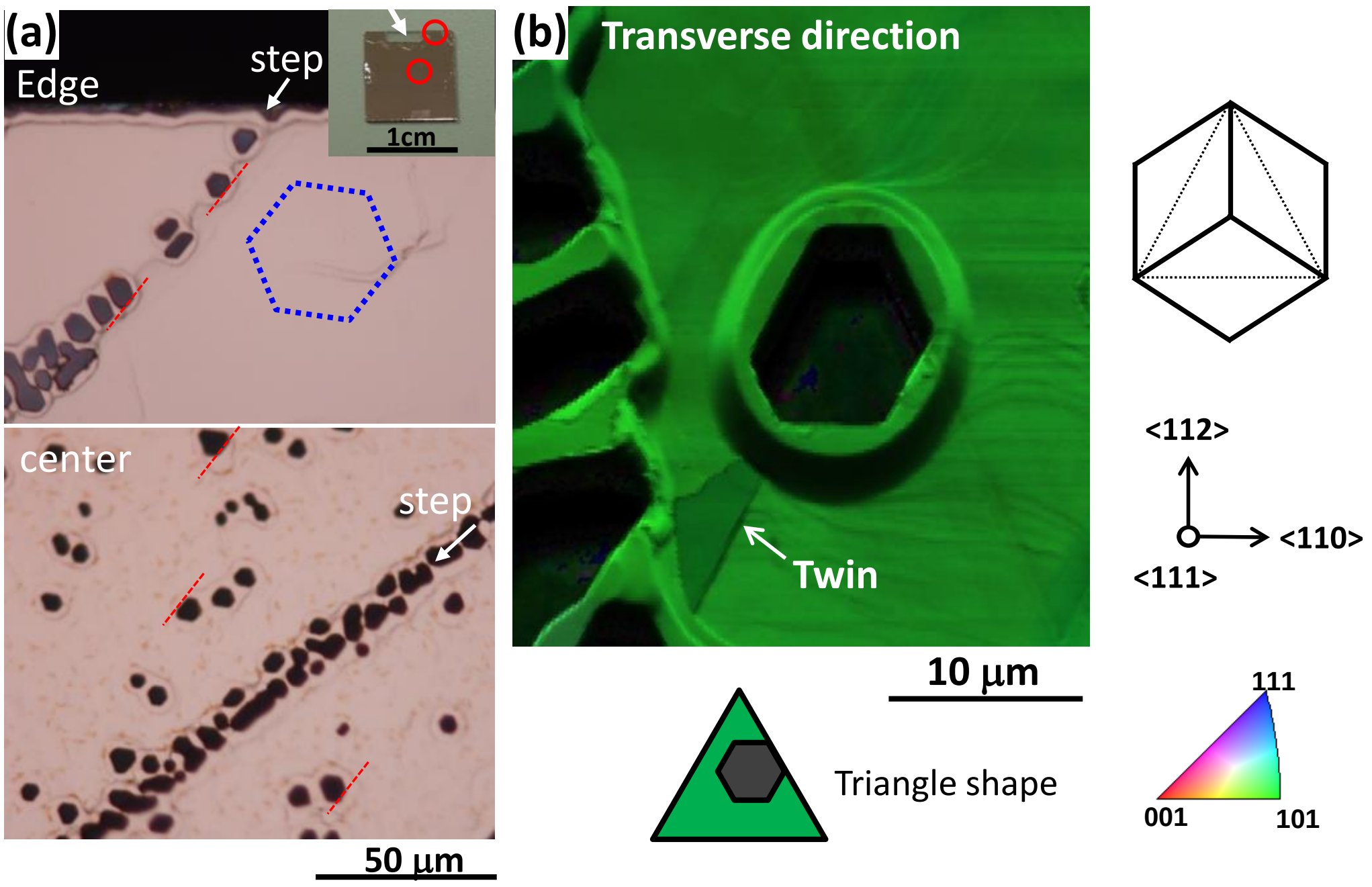

**Fig. 3** (a) Optical micrographs at the edge and center of the Cu/mica substrate (red circles in the inset). (b) EBSP map for the hexagonal void and the triangle shape. The EBSP map is colored for the transverse direction, not for the normal direction, because the triangular shape is distinct in green color but not in red color. The relationship between the hexagonal void and triangular shape is illustrated at the bottom.

this EBSP map using the inverse pole figure triangle (shown at the bottom right corner). Although the Cu film is macroscopically single crystal, the twins are often observed in the EBSP map, as indicated by a white arrow. The twin formed by the 60° rotation on the <111> axis is not distinctly perceived in an observation from the <111> direction. Therefore, the twin can be identified from the EBSP map by tilting the sample slightly from the <111> direction. The important information in **Fig. 3(b)** is that the triangular shape is observed. To help detecting the triangle shape, the schematic illustration is also shown. This triangle shape is consistent with that observed in the AFM image in **Fig. 1(b)** and has the specific orientation relation with the hexagonal shape of the void. The orientations of hexagonal and triangular shapes are revealed based on the orientation of Cu, as shown by the cube on the right side of the figure.

**Figure 4** is a schematic of orientation relationships among mica(001), Cu(111) and graphene. The surface structure of mica is referenced from the text book [18], and the orientation relation between graphene and Cu(111) is based on the previous reports [5,6]. Two types of stable positions for graphene on Cu(111) are illustrated because its position has not been uniquely determined because of the small lattice mismatch. Regardless of these two types of stable positions, the macroscopic orientations of graphene, e.g., zigzag and armchair, can be recognized based on the hexagonal void shape of Cu (the dotted hexagonal shape in the figure). Although the voids in **Fig. 3** were intentionally formed throughout the Cu/mica substrate for the annealing time longer than 1 hour, they were also observed near the pin that is used to hold the substrate even for the present CVD growth condition. The position of the pin (no Cu film region) is shown by a white arrow in the inset of **Fig. 3(a)**. This can be used as the mark to recognize the graphene orientation. Considering both the well-identified orientation of graphene and the reduced macroscopic surface roughness, graphene/Cu(111)/mica(001) can be used as the graphene substrate for the top-gate insulator research on graphene.

### 3-2. CVD growth of graphene and its electrical properties

During the graphene growth on Cu/mica in the $CH_4/Ar/H_2$ gas flow, $H_2$ gas plays multiple roles, such as the cleaning and chemical reduction of the Cu surface, edge reconstruction, etching of graphene domain and the removal of surface-adsorbed C atoms [19-21]. **Figure 5** shows the coverage ratio and $I_D/I_G$ ratio of SLG as a function of the $H_2$ gas flow rate. The gas flow rates of $CH_4$/Ar, the total pressure and the growth period were kept at 20/100 sccm, ~1000 Pa and 15 min, respectively. The SLG coverage ratio is

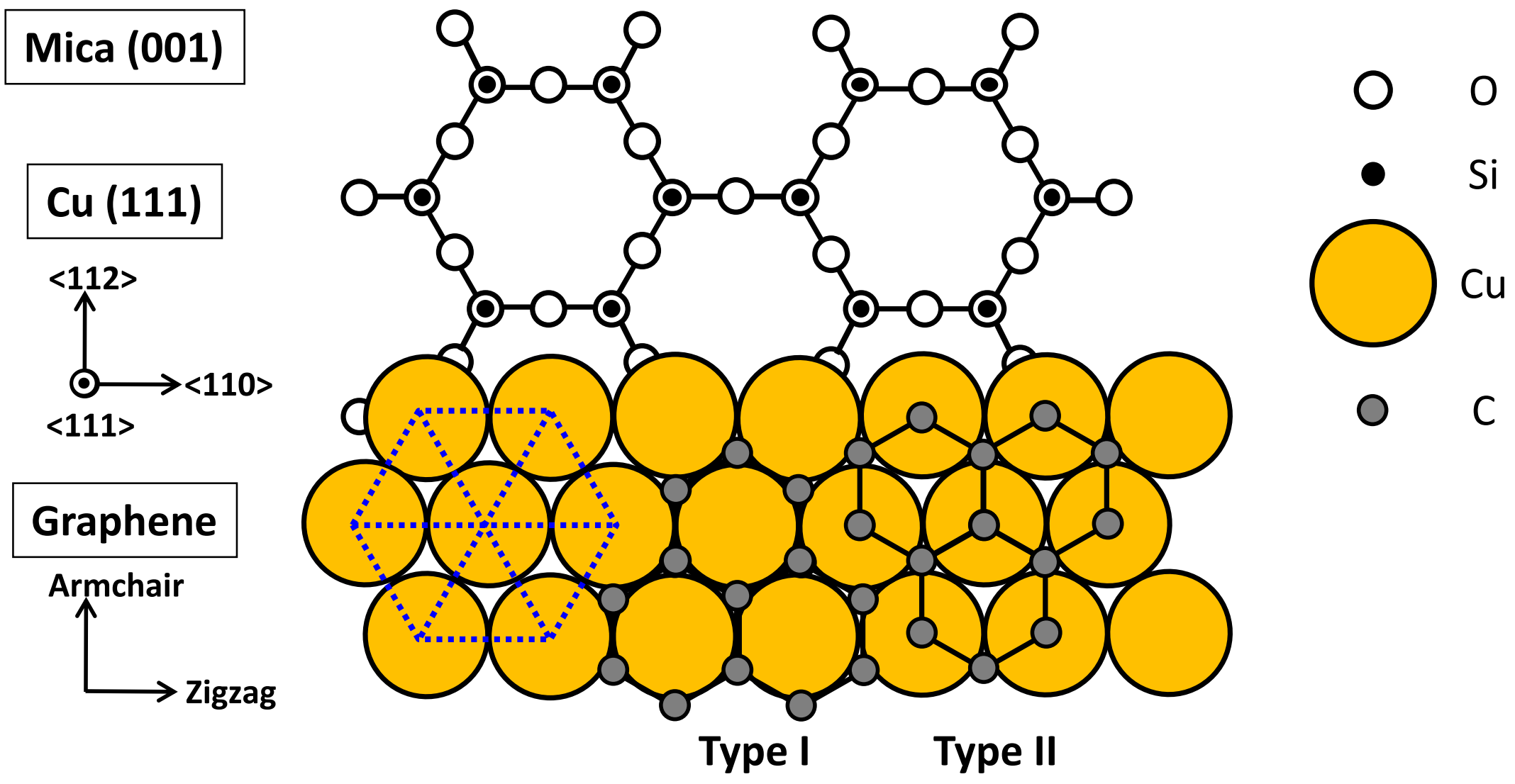

**Fig. 4** Schematic of orientation relationship among mica(001), Cu(111) and graphene.

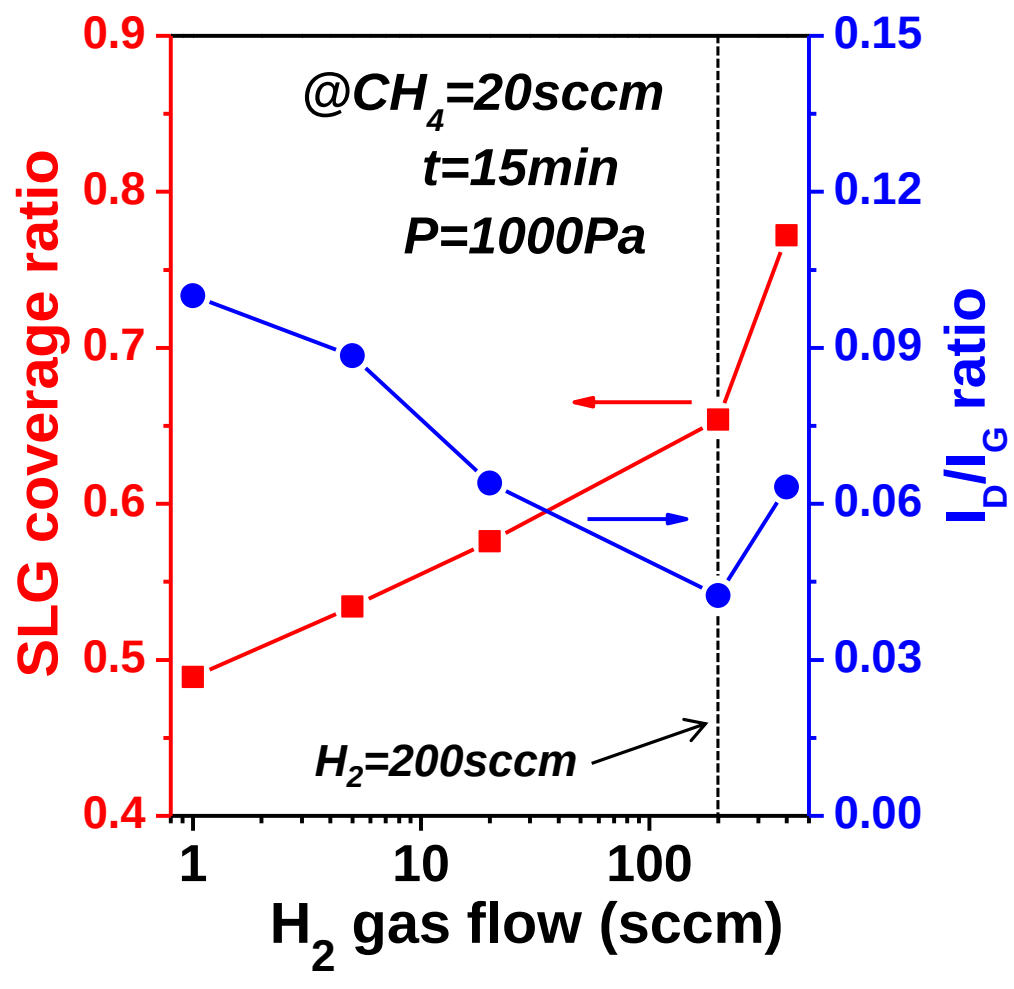

**Fig. 5** SLG coverage ratio and $I_D/I_G$ ratio as a function of the $H_2$ flow rate. In the growth experiment, the flow rate of $CH_4$ and Ar are fixed at 20 and 100 sccm, and the total pressure is ~1000 Pa.

defined as $A_{SLG}/(A_{SLG}+A_{MLG})$, where $A$ and MLG denote the area and multi-layer graphene, respectively. The SLG coverage ratio and $I_D/I_G$ ratio were measured for graphene transferred onto an $SiO_2$/Si wafer. The SLG coverage ratio increases monotonically with an increasing $H_2$ gas flow rate, while the $I_D/I_G$ ratio shows the minimum value at an $H_2$ gas flow rate of 200 sccm. The nucleation rate of graphene is reduced when the $H_2/CH_4$ ratio exceeds unity because the coverage ratio of $H_2/CH_4$ on the Cu surface is reversed and the $CH_4$ decomposition is reduced by the high $H_2$ flow rate [19]. The growth rate of graphene is kept slow in this condition, and the graphene quality is

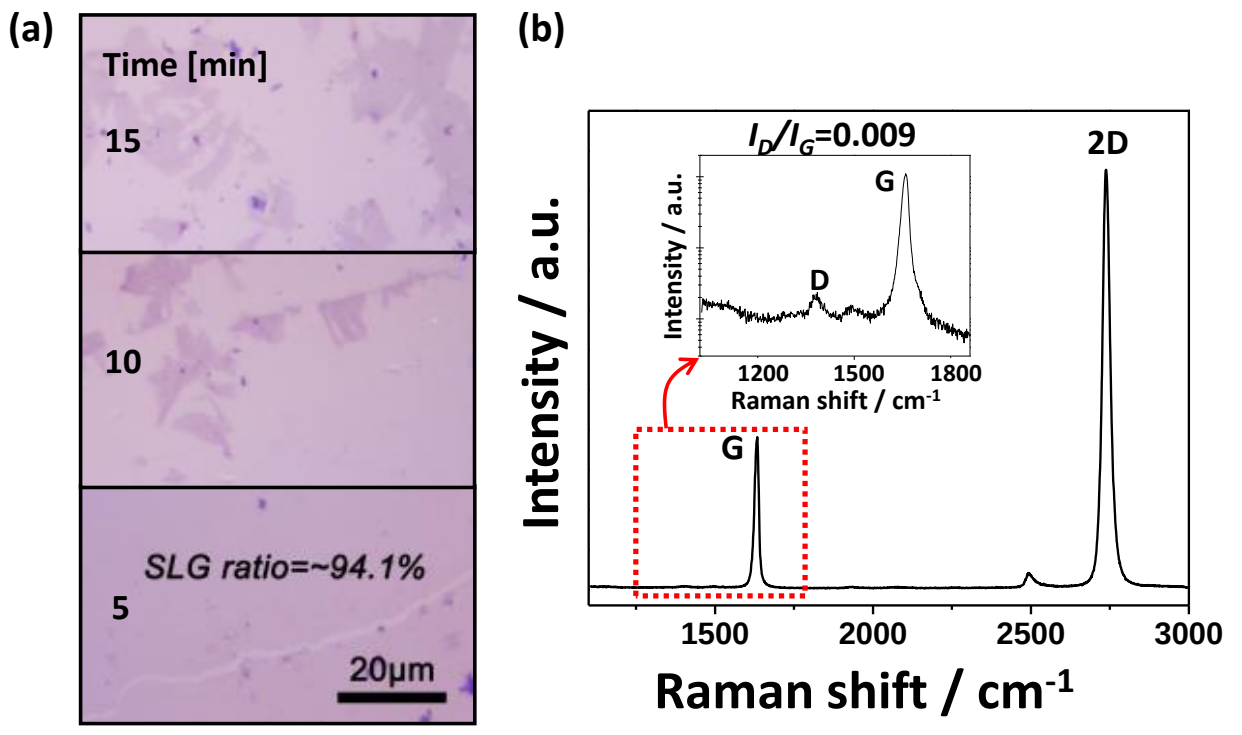

**Fig. 6** (a) Optical micrographs of graphene transferred onto $SiO_2$/Si with different growth times (from 15 to 5 min). The flow rates of $H_2$, $CH_4$ and Ar are fixed at 20, 1 and 979 sccm at the ambient pressure, respectively. (b) Raman spectroscopy of graphene transferred on $SiO_2$/Si for a growth time of 5 min. The vertical axis in the inset is converted to log scale.

increased, as indicated by the $I_D/I_G$ ratio. However, when the $H_2$ flow rate exceeds 200 sccm ($H_2/CH_4$ flow ratio>10), the increase in the $I_D/I_G$ ratio indicates the introduction of defects such as the C-H bonding in graphene [19,20]. These results indicate that there is an optimal $H_2$ gas flow rate under the fixed $CH_4$ flow rate and fixed total pressure. Moreover, the SLG coverage can be increased by up to ~95 % by reducing the growth time from 15 to 7.5 min ($H_2/CH_4$/Ar=200/20/100 sccm).

To decrease the $I_D/I_G$ ratio, the total pressure was increased to the ambient pressure because the stable decomposition of $CH_4$ on the Cu surface was achieved by suppressing the evaporation and agglomeration of Cu [10-13]. The gas flow rates of $H_2/CH_4$/Ar and substrate temperature were fixed at 20/1/979 sccm and 1000 °C, respectively. Then, the growth time was decreased from 15 to 5 min. **Figure 6(a)** shows the optical micrographs of graphene transferred onto $SiO_2$/Si with different growth times, ranging from 15 min to 5 min. A typical Raman spectrum of graphene with the growth time of 5 min is shown in **Fig. 6(b)**. It is evident that the film is the single-layer graphene [14] and that the $I_D/I_G$ ratio of graphene has significantly improved to 0.009 from 0.033 for the low-pressure case.

**Figure 7** summarizes the graphene growth on Cu(111) film and Cu polycrystalline foil using the difference in total pressure. There are two dominant factors which result in the reduction of the nucleation rate and growth rate: the lattice mismatch between Cu and graphene and the evaporation of Cu. For graphene growth on Cu polycrystalline foil at a low total pressure, both the nucleation and growth rates are lowered by the lattice mismatch and the evaporation of Cu. On the other hand, for graphene growth on the Cu(111) film at the ambient pressure, both the nucleation and growth rates are the fastest among the three cases. Controlling these two factors is thus critical for growing high-quality graphene.

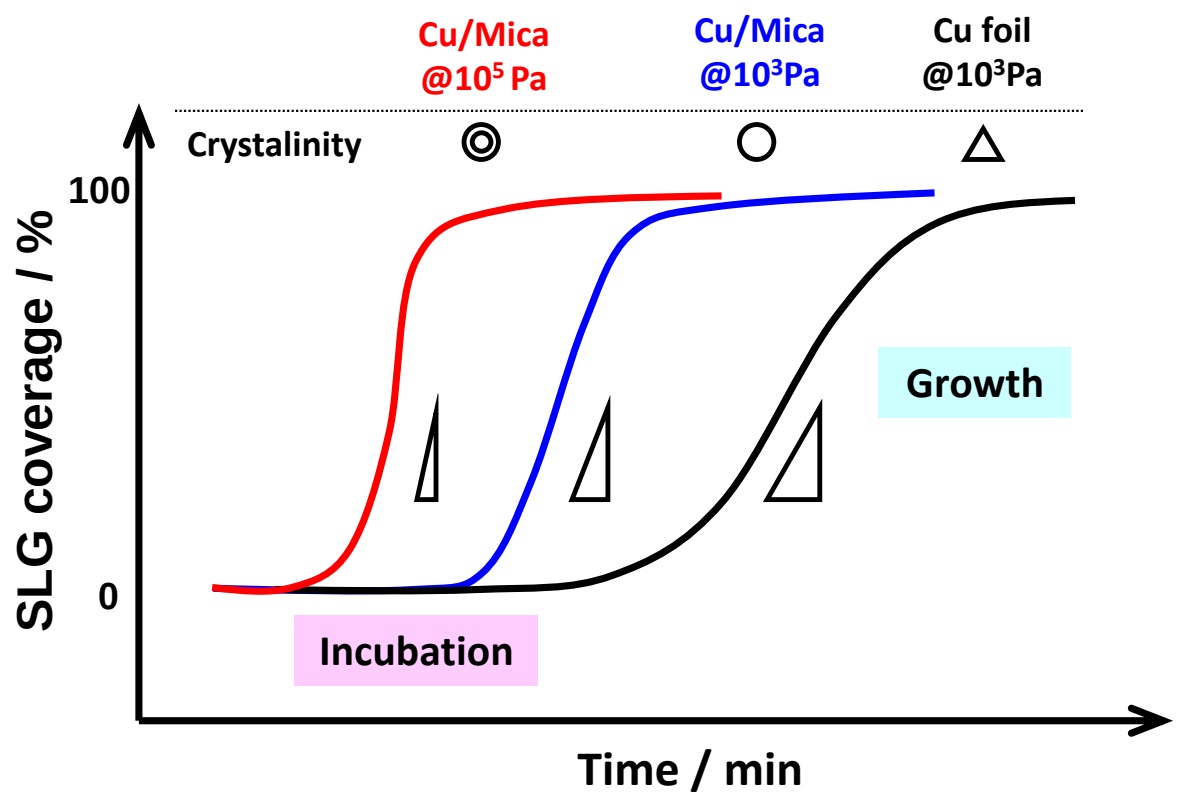

**Fig. 7** Schematic summary of the graphene growth for different growth conditions.

Finally, electrical properties of graphene transferred onto $SiO_2$/Si were characterized using the four-probe measurement. **Figure 8** shows (a) the optical micrograph of the typical graphene field effect transistor, (b) the sheet resistivity as a function of carrier density and (c) the mobility extracted at carrier density of $1\times10^{12}$ $cm^2$ in graphene obtained in **Fig. 6(b)**. Compared with the data reported for the mechanical exfoliation of Kish graphite [15], there are still spaces to be improved for CVD growth of graphene. For example, although graphene is hetero-epitaxially grown on Cu(111)/mica(001), small-angle grain boundaries caused by the growth from different nucleation sites may exist within the graphene channel of ~10 μm in length. However, mobility as high as ~4500 $cm^2V^{-1}s^{-1}$ was achieved by reducing Cu evaporation during CVD growth. This result indicates that graphene grown under ambient pressure is highly recommendable for the graphene substrate.

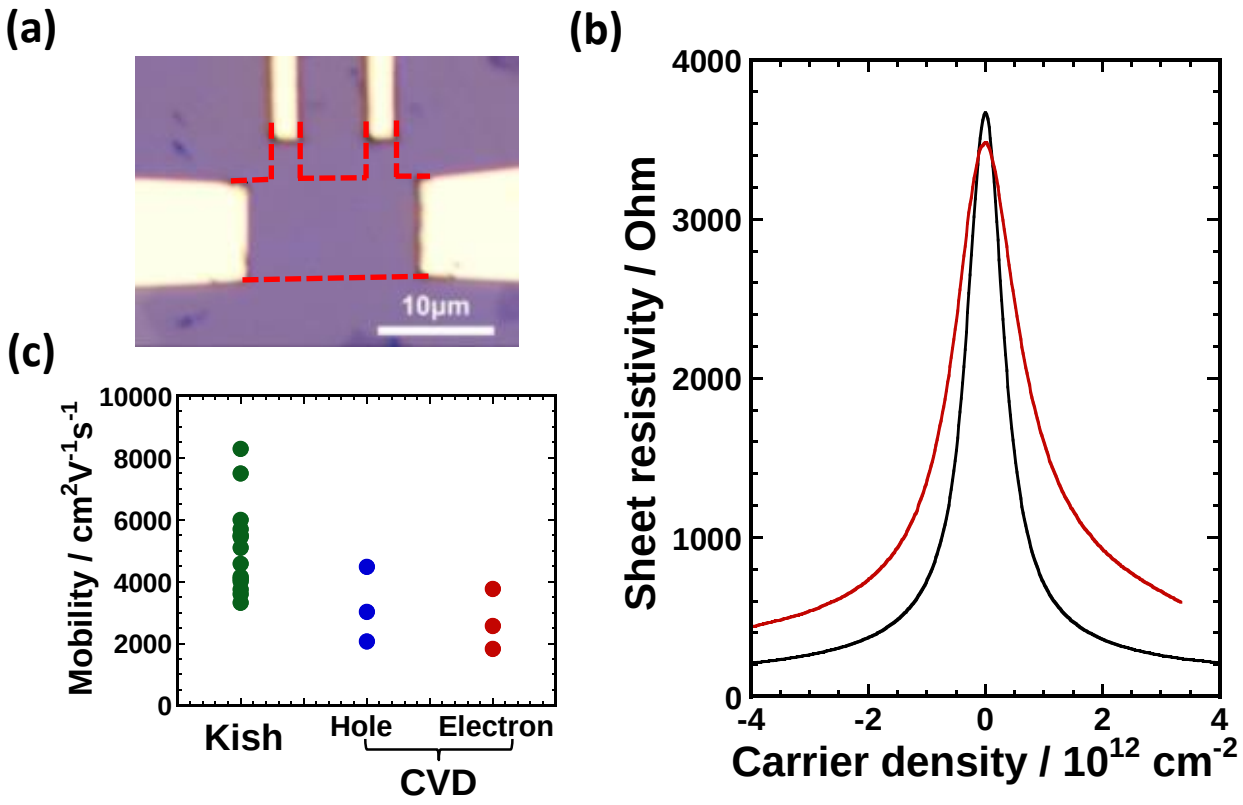

**Fig. 8** (a) Optical micrograph of a typical graphene field effect transistor. The graphene channel was etched into the four-probe shape by $O_2$ plasma (red dotted line). (b) Mobility extracted at the carrier density of $1\times10^{12}$ $cm^{-2}$. (c) Sheet resistivity as a function of carrier density for graphene transferred mechanically from Kish graphite and CVD graphene in Fig. 6(b).

## 4. Summary

We have studied the hetero-epitaxial graphene growth on Cu(111)/mica(001) by chemical vapor deposition to assess its applicability as the graphene substrate for the top-gate insulator research on graphene. The AFM image indicates that the average smooth area is ~100 $\mu m^2$. The orientation of graphene in the graphene/Cu/mica substrate is identified from the EBSP orientation mapping for the hexagonal void morphology of Cu, based on previous reports on the orientation relation between graphene and Cu(111). Furthermore, the crystalline quality of graphene is improved by reducing the Cu evaporation in the ambient-pressure CVD growth. We demonstrate the high mobility of ~4500 $cm^2V^{-1}s^{-1}$ in graphene transferred on the $SiO_2$/Si substrate. These results suggest that the present graphene/Cu/mica structure can be used as the graphene substrate for the top-gate insulator research on graphene.

**Acknowledgements**

The authors thank Dr. Chiashi, The University of Tokyo, and Dr. T. Fujii, Fuji Electric Co., Ltd., for their useful comments and suggestions. J. L. Q. acknowledges the financial support from the National Natural Science Foundation of China under Grant No. 51105108. K. N. acknowledges the financial support from a Grant-in-Aid for Scientific Research on Innovative Areas and for Challenging Exploratory Research from the Ministry of Education, Culture, Sports, Science and Technology, Japan.